\documentclass[11pt]{article}

\usepackage{graphicx,comment}
\usepackage[linesnumbered,vlined]{algorithm2e}
\RestyleAlgo{boxruled}

\usepackage{amsmath}
\usepackage{amsthm}
\usepackage{amsfonts, amssymb}
\usepackage{cite}
\usepackage{fullpage}

\newtheorem{open}{Open Problem}[section]

\newcommand{\ccfull}{\ensuremath{\mathsf{CONGESTED~CLIQUE}}\xspace}
\newcommand{\clique}{\ensuremath{\mathsf{CLIQUE}}\xspace}
\newcommand{\congest}{\ensuremath{\mathsf{CONGEST}}\xspace}
\newcommand{\bcongest}{\ensuremath{\mathsf{BROADCAST~CONGEST}}\xspace}
\newcommand{\local}{\ensuremath{\mathsf{LOCAL}}\xspace}
\newcommand{\bcc}{\ensuremath{\mathsf{BROADCAST~CONGESTED~CLIQUE}}\xspace}

\def\polylog{\operatorname{polylog}}

\usepackage{color}

\begin{document}
\title{Distributed Subgraph Finding: Progress and Challenges}
\author{Keren Censor-Hillel\\ Department of Computer Science, Technion\\ \texttt{ckeren@cs.technion.ac.il}}
\date{}
\maketitle

\begin{abstract}
This is a survey of the exciting recent progress made in understanding the complexity of distributed subgraph finding problems. It overviews the results and techniques for assorted variants of subgraph finding problems in various models of distributed computing, and states intriguing open questions. This version contains some updates over the ICALP 2021 version, and I will try to keep updating it as additional progress is made.

~\\Revisions (most recent first):
\begin{itemize}
\item{September 2025:} 
Added new lower bound for triangle detection in \congest.\\
Added new results on deterministic cycle detection in \congest.
\item{December 2024:} Added new results on deterministic clique listing in \congest.\\
Added new results on cycle detection in \congest.\\
Added new results on cycle detection in quantum \congest.\\
Added new results on parametrized cycle detection in \clique.\\
Added new results on the distance to triangle freeness.
\item{December 2022:} Corrected the variant of triangle finding studied with respect to the number of triangles in the graph -- from listing to detection.\\
Corrected a typo in running time of sparse matrix multiplication.
\item{November 2022:} Added new results on deterministic clique listing and quantum algorithms.
\end{itemize}
\end{abstract}

\section{Introduction}
Distributed subgraph finding is motivated by considering a user in a network whose connections are unknown to its users. Typical examples could be friends or followers in social networks, or computing devices in large networks. A prime question is what is the local structure of the network, e.g., do two connected users share common connections, or are they part of a slightly larger cycle or clique? While some social networks provide the user with information about common connections they have with their connections, the general fundamental question that arises is that of finding small subgraphs in such distributed settings. This type of questions also emerges when processing huge graphs by distributed systems. For example, Hirvonen, Rybicki, Schmid, and Suomela~\cite{HirvonenRSS17} study distributed algorithms for finding large cuts in triangle-free graphs, and Pettie and Su~\cite{PettieS13}  study coloring in triangle-free graphs.

This article surveys the state-of-the-art in distributed subgraph finding. We focus on synchronous settings, in which the main complexity measure is the number of communication rounds that is required in order to find a subgraph $H$. We will elaborate on the computational models and on the possible interpretations of what it means to \emph{find} a subgraph, but let us start with a warm-up.

~\\\textbf{Warm-up: locality.} The first simple observation is that in a synchronous network with no additional restrictions, the round complexity of finding a subgraph $H$ by the devices in the network is $\Theta(k)$, where $k$ is the diameter of $H$. The reason for this is that in a single round of communication, all nodes of the network can learn the neighbors of their neighbors,\footnote{This assumes that all nodes begin with knowledge of their neighbors. Removing this assumption incurs another round of communication to this argument.} and by induction, within $t$ rounds each node can learn the topology within its $t$-neighborhood.

To be more precise, what we get is that within $O(k)$ rounds, each node is able to \emph{list} all the instances of $H$ to which it belongs. This is the most powerful variant of subgraph finding.

The clear drawback of the above example is that it sweeps the under the carpet the complexity of sending potentially very large messages in a single round. Therefore, while the above classic \local model~\cite{Linial92} is suitable for studying many other tasks, it misrepresents the complexity of subgraph finding. We will thus focus mainly on two distributed settings that capture limited bandwidth, and in which the complexity of some subgraph finding problems is related. Additional settings will be discussed towards the end of the survey.

~\\\textbf{Computational models.}
The standard model that imposes restrictions on the bandwidth over the above setting is the \congest model~\cite{PelegBook}. This is a synchronous model, in which each of $n$ nodes can send a message to each of its neighbors in every round, where the size of messages is bounded by $O(\log{n})$ bits. This choice of this bound on the size of messages allows sending a node identifier within a single message. Typically, the graph that is processed is the graph that underlies the communication network.

Another important model is the \ccfull model~\cite{LotkerPPP05}, which we will abbreviate as the \clique model in this document, in which the $n$ nodes are part of a fully connected network, with the same message bound of $O(\log{n})$ bits as in the \congest model. Here, the input graph is an arbitrary graph $G$ on $n$ nodes, which is typically assigned through a bijection to the nodes of the \clique, such that each node receives as input the edges in $G$ that are adjacent to its assigned node.

~\\\textbf{Subgraph finding.} The extreme case mentioned above, where each node in the system lists all instances of $H$ to which it belongs is referred to as \emph{membership listing} (or sometimes \emph{local listing} or \emph{local enumeration}). In settings with bounded bandwidth, this is typically a difficult variant that is known to require many rounds of computation for many subgraphs (see, e.g., discussion about triangle finding in Section~\ref{section:triangles}). A weaker variant is that of \emph{listing} or \emph{enumeration}, in which it is required that every instance of the subgraph $H$ is output by some node, but not necessarily a node which belongs to that instance.

Apart from these two listing variants, it is in many cases important to merely \emph{detect} whether the input contains a copy of $H$ or not. The standard formalization of distributed detection is that if there is no copy of $H$ then all nodes output \texttt{false}, and otherwise at least one node outputs \texttt{true}. The reason for not demanding a unanimous output is to avoid incurring an overhead just for propagating the information of existence of $H$, for example if a graph contains a single copy of $H$ and has many nodes that are very far from that instance.

Similarly to the case of listing, the detection problem also has a membership variant, which requires each node to output \texttt{true} or \texttt{false} based on whether it is a part of a copy of $H$ or not.

~\\\textbf{Outline.} Section~\ref{section:triangles} overviews the case in which $H$ is a triangle. It covers both the \clique and the \congest models. Sections~\ref{sec:clique} and~\ref{sec:cycles} address larger cliques and larger cycles in both models, respectively. Finally, Section~\ref{sec:additional} briefly mentions additional subgraphs and additional settings.

\section{Triangle Finding}
\label{section:triangles}
A na\"ive simulation of the warm-up algorithm for membership listing in the \congest or \clique models, in which all neighbors are sent to each other neighbor one by one, gives a trivial $O(\Delta)$-round algorithm for triangle membership listing, where $\Delta$ is the maximum degree in the graph. However, it is possible to do better, as we overview in this section.

\subsection{Triangle Finding in the \clique Model}
\label{section:triangles-clique}

We begin with the \clique model.

~\\\textbf{Triangle listing in the \clique model.} The first non-trivial algorithm for triangle finding is due to Dolev, Lenzen, and Peled~\cite{Dolev+DISC12}. This is a deterministic triangle listing algorithm for the \clique model, which has a complexity of $O(n^{1/3}/\log{n})$ rounds. The simplicity of this algorithm turned out to be a huge advantage for later additional results, as we will see. The algorithm works as follows: The vertices of the graph are partitioned into $n^{1/3}$ subsets $S_1,\dots,S_{n^{1/3}}$, each of $n^{2/3}$ nodes. Each of the $n$ nodes receives a different tuple of three of these subsets. A node that receives $S_{i_1},S_{i_2},S_{i_3}$ for indices $1\leq i_1,i_2,i_3 \leq n^{1/3}$ (that are not necessarily different) collects all edges with one endpoint in one of the three subsets and one endpoint in another, that is, this node collects all edges in $E(S_{i_1},S_{i_2})\cup E(S_{i_1},S_{i_3}) \cup E(S_{i_2},S_{i_3})$, and reports all triangles that it finds. It is straightforward to see that all triangles are listed by this algorithm since the number of 3-tuples of subsets is $n$ and so each is handled by some node.

The round complexity of the algorithm follows by proving that each node needs to send and receive $O(n^{4/3})$ edges in total, which are to and from locations that are known to all nodes (we will discuss this knowledge property later), since the partition to subsets is hardcoded and so it is known to all nodes. Sending: Take a node $v$ and assume that it is in the subset $S_i$. There can be at most $n^{2/3}$ edges between $v$ and nodes in $S_j$ and these edges need to be sent to all nodes that have $S_i$ and $S_j$ in their 3-tuple. Since there are $n^{1/3}$ such 3-tuples, these $n^{2/3}$ edges need to be sent to $n^{1/3}$ nodes. Repeating this for all $n^{1/3}$ possibilities for $j$ gives a total of $n^{2/3+1/3+1/3}=n^{4/3}$ edges that $v$ has to send. Receiving: Each node needs to learn 3 subsets of edges, each containing at most $n^{2/3}\cdot n^{2/3} = n^{4/3}$ edges. To conclude the complexity analysis, one can use the simple claim that~\cite{Dolev+DISC12} proves, which states that a routing task in which each node needs to send and receive $n$ messages in a known pattern can be done in 2 rounds. This means that the $O(n^{4/3})$ sent and received messages per node are divided by $n$, yielding a complexity of $O(n^{1/3})$ rounds. Noticing that the partition and routing are fixed, one can refrain from sending actual edge identifiers and replace them with a bit mask, which saves a logarithmic factor and results in a complexity of $O(n^{1/3}/\log{n})$ rounds.

This complexity turns out to be optimal. The first lower bound for this task was of  $\Omega(n^{1/3}/\log^3{n})$ rounds given by Pandurangan, Robinson and Scquizzato~\cite{Pandurangan+SPAA18}, and it was followed by a tight lower bound of $\Omega(n^{1/3}/\log{n})$ rounds given by Izumi and Le Gall~\cite{Izumi+PODC17}. The lower bound is proven using an information-theoretic argument, which bounds by $\Omega(n^{4/3})$ the entropy of the transcript that a certain node sees on a random graph in which each edge appears independently at random with probability $1/2$. As the entropy of the transcript is a lower bound on its length, and since each node can receive at most $O(n\log{n})$ bits per round, the lower bound on the round complexity follows. This approach also allowed~\cite{Izumi+PODC17} to obtain a lower bound of $\Omega(n/\log{n})$ rounds for \emph{local} listing of triangles, in which each node needs to output a list of all the triangles which include it.

~\\\textbf{Triangle detection in the \clique model.} The approach of~\cite{Dolev+DISC12} inherently lists all triangles. A natural question is whether the decision problem of triangle detection is easier than listing, for which the answer turns out to be affirmative. Censor-Hillel, Kaski, Korhonen, Lenzen, Paz, and Suomela~\cite{Censor-Hillel+DC19} show how to perform matrix multiplication over a ring in the \clique model within $O(n^{0.158})$ rounds. More precisely, the complexity is $O(n^{1-2/\omega})$ rounds, where $\omega$ is the matrix multiplication exponent, currently known to be bounded by $2.3728596$ due to Alman and Vassilevska-Williams~\cite{AlmanVW21}. Ring matrix multiplication directly carries over to triangle detection with the same complexity.
The main approach of~\cite{Censor-Hillel+DC19} is to simulate matrix multiplication algorithms for parallel settings in the \clique model. This includes the so-called parallel 3D matrix multiplication, as well as bilinear Strassen-like algorithms.

Elaborating upon the matrix multiplication algorithms is beyond the scope of this survey. Yet, we must mention a crucial component that underlies these \clique algorithms, as well as numerous additional \clique algorithms for various tasks, which is the ingenious routing technique of Lenzen~\cite{Lenzen13}. In a nutshell, this technique provides a way to route in $O(1)$ rounds any set of messages in which each node needs to send and receive at most $n$ messages.

The above complexity of $O(n^{0.158})$ rounds for triangle detection in the \clique model is the current state-of-the-art.   For reasons that will be explained shortly, unlike the $\Omega(n^{1/3})$ lower bound for triangle listing in this model, there is no known lower bound for triangle detection. We thus have the following major open problem in distributed triangle finding.

\begin{open}
\label{open:trianlge-detection-clique}
What is the complexity of triangle detection in the \clique model? In particular, is there an algorithm that is faster than $O(n^{0.158})$ rounds? Can any lower bound be proven?
\end{open}

The second part of the above question which asks for a lower bound for triangle detection in the \clique model is considered hard: Very roughly speaking, Drucker, Kuhn, and Oshman~\cite{Drucker+PODC14} show that the \clique model is strong enough to simulate certain circuits, which implies that for a wide range of problems, any super-constant lower bound in the \clique model would result in a major breakthrough in circuit complexity. It is noteworthy that the problem of triangle listing does not fall into this category of problems due to its large outputs, which explains why this does not contradict the aforementioned lower bound of $\Omega(n^{1/3})$ rounds for triangle listing.

\sloppy{
In addition to the aforementioned algorithm of~\cite{Dolev+DISC12}, the paper also gives a triangle-detection algorithm whose complexity improves upon the above in the case that the graph contains \emph{many} triangles. Specifically, they provide a randomized algorithm that completes in $O(n^{1/3}/(t^{2/3}+1)+1)$ rounds in expectation, where $t$ is the number of triangles in the graph, and in $O(\min\{n^{1/3} \log^{2/3}{n}/(t^{2/3} + 1), n^{1/3}\})$ rounds with high probability. Later, Censor-Hillel, Even, and Vassilevska Williams~\cite{Censor-HillelEW24} improved this complexity to $\tilde{O}(n^{0.1567}/(t^{0.393}+1))$ rounds. This also holds for directed graphs. The main tool in obtaining the above is a technique for computing the products of many small random square matrices.
}

For the broadcast version of this model, a.k.a. the \bcc model, in which the messages sent by a node in a certain round must all be identical, an $\Omega(n/e^{O(\sqrt{\log{n}})}\log{n})$-round lower bound on deterministic triangle detection is given by~\cite{Drucker+PODC14}, through a reduction from 3-party number-on-forehead set disjointness.

~\\\textbf{Triangle listing in sparse graphs in the \clique model.}
The first algorithms for triangle finding in sparse graphs were given by Dolev, Lenzen, and Peled~\cite{Dolev+DISC12}. One algorithm has a round complexity of $O(\Delta^2/n+1)$ and another has a round complexity of $O(A^2/n+\log_{2+n/A^2}{n})$, where $A$ is the arboricity of the graph.\footnote{The arboricity of a graph is the minimal number of forests that contain all of its edges. While bounded by the maximum degree $\Delta$, the arboricity can in some cases be much smaller, e.g., a star has a linear maximum degree but its arboricity is 1.} The latter implies a round complexity of ${O}(m^2/n^{3})$, in terms of the number of edges, $m$.

Pandurangan, Robinson, and Scquizzato~\cite{Pandurangan+SPAA18} give a randomized triangle listing algorithm that completes within  $\tilde{O}(m/n^{5/3} + 1)$ rounds, w.h.p. At the heart of the algorithm lies the partitioning approach of~\cite{Dolev+DISC12}, but more is needed in order to exploit sparsity. In~\cite{Pandurangan+SPAA18}, the partition is random, and the routing of edges to the nodes to which they are assigned is done in a randomized load balanced manner. This approach is what handles load balancing of the information that needs to be routed in the system in a way which is sensitive to the sparsity of the graph, rather than optimizing only for the worst case.
Moreover, the aforementioned $\tilde{\Omega}(n^{1/3})$ lower bound of ~\cite{Pandurangan+SPAA18} for triangle listing in general graphs follows from a $\tilde{\Omega}(m/n^{5/3})$ lower bound for graphs with $m$ edges.\footnote{In fact, the results of~\cite{Pandurangan+SPAA18} hold for the $k$-machine model (see Klauck, Nanongkai, Pandurangan, and Robinson~\cite{KlauckNP015}), which is similar to the \clique model, but has $k$ nodes rather than $n$. For a general $k$, the complexity of the algorithm is $\tilde{O}(m/n^{5/3} + n/k^{4/3})$, and the lower bound is $\tilde{\Omega}(m/k^{5/3})$.}

Censor-Hillel, Leitersdorf, and Turner~\cite{Censor+TCS20} obtained an algorithm for sparse graphs, with a  complexity of ${O}(m/n^{5/3} + 1)$ which is similar to, but slightly improves upon, the complexity  of the algorithm of~\cite{Pandurangan+SPAA18}, by polylogarithmic factors. The algorithm of~\cite{Censor+TCS20} is deterministic and is based on an algorithm for sparse matrix multiplication, which completes in
%$O(nz(S)^{1/3}nz(T)^{1/3}/n+1)$
$O((\rho_S\rho_T/n)^{1/3}+1)$  rounds, where $S$ and $T$ are the input matrices and $\rho_A$ is the average number of non-zero elements per row of a matrix $A$ (i.e., it is the number of non-zero elements of $A$, divided by $n$).\footnote{The sparse matrix multiplication algorithm of~\cite{Censor+TCS20} can also be converted to the $k$-machine model, in which it has a complexity of $O(n^{4/3}(\rho_S\rho_T)^{1/3}/k^{5/3}+1)$ rounds.} %and $nz(A)$ is the number of non-zero elements in a matrix $A$.\footnote{The algorithm of~\cite{Censor+TCS20} can also be converted to the $k$-machine model, in which it has a complexity of $O(n^{2/3}nz(S)^{1/3}nz(T)^{1/3}/k^{5/3}+1)$ rounds.} \todo{the approach, compared to general semiring.}
While the aforementioned semi-ring matrix multiplication algorithm of~\cite{Censor-Hillel+DC19} assigns the $n^3$ element-wise multiplication to the nodes in an optimal manner, it applies to the worst case. The general approach used by\cite{Censor+TCS20} for sparse matrix multiplication is to assign the element-wise multiplications to nodes in a way which is optimal given the input sparsity. The way this is done is by load balancing the number of non-zero elements that each node needs to send and receive in order for all element-wise multiplications to be computed.

~\\\textbf{Notes on matrix multiplication in the \clique model.} The first algorithm for matrix multiplication in this model is due to Drucker, Kuhn, and Oshman~\cite{Drucker+PODC14}, who showed a randomized complexity of $O(n^{\omega-2})\approx O(n^{0.373})$ rounds with high probability, for semirings. A deterministic $O(n^{1/3})$-round algorithm for matrix multiplication over a semiring is given in~\cite{Censor-Hillel+DC19}.

Censor-Hillel, Dory, Korhonen, and Leitersdorf~\cite{CHDKL20} provide two additional sparse matrix multiplication algorithms, used for distance computations. Compared to~\cite{Censor+TCS20}, these algorithms also benefit from sparsity of the output matrix $P=ST$. More concretely, their first algorithm completes in
$O((\rho_S\rho_T\rho_P)^{1/3}/n^{2/3}+1)$  rounds. This matches the complexity algorithm of~\cite{Censor+TCS20} for a general $P$, but improves upon it for sparse output matrices (recall that the multiplication of two sparse matrices need not be sparse in general). The second algorithm of~\cite{CHDKL20} pays an additive $\log{n}$ over the first one, i.e., has a round complexity of $O((\rho_S\rho_T\rho)^{1/3}/n^{2/3}+\log{n})$, but here $\rho_P$ is replaced by $\rho$, which stands for the number of elements per row that are needed. That is, there is no need to compute any element in the output matrix $P$ that is not among the $\rho$ smallest ones in its row (given an appropriate definition of the total order on matrix elements). In other words, the complexity of this algorithm does not depend on the sparsity of $P=ST$, but on some sparsity parameter that is given as input.

Additional algebraic algorithms in the \clique model with important applications were given by Le Gall~\cite{LeGall16}. These include fast algorithms for rectangular matrix multiplications, as well as for multiple instances of matrix multiplication. An approach for multiple small instances that are in some sense balanced (such as random instances) was given by Censor-Hillel, Even, and Vassilevska Williams~\cite{Censor-HillelEW24}.

As explained earlier, we do not expect a lower bound for matrix multiplication in the \clique model. However, it is shown in~\cite{Censor-Hillel+DC19} that a near-linear number of rounds is required in the  \bcc model.

Finally, we mention that matrix multiplication based algorithms also give results for \emph{counting} for some small subgraphs~\cite{Censor-Hillel+DC19, Censor+TCS20}.

\subsection{Triangle Finding in the \congest Model.}
\label{section:triangles-congest}

We now move to triangle finding problems in the \congest model.

The first breakthroughs in this model are the first non-trivial (sublinears) algorithms due to Izumi and Le Gall~\cite{Izumi+PODC17}. These are randomized algorithms for triangle listing and detection in $O(n^{3/4}\log{n})$ and $O(n^{2/3}\log^{2/3}{n})$ rounds w.h.p., respectively. The approach of these algorithms is to split the task of finding triangles into two parts: one which looks for triangles that are $\epsilon$-heavy and another which looks for other triangles, where an $\epsilon$-heavy triangle is one in which at least one of its edges appears in at least $n^{\epsilon}$ triangles. Very roughly speaking, heavy triangles can be detected within $O(n^{1-\epsilon})$ rounds by randomly sampling which edges to send, and can be listed in $O(n^{1-\epsilon/2})$ rounds by randomly hashing the edges sent to different neighbors. A more involved argument shows that non-heavy triangles can be listed in $O(n^{1-\epsilon}+n^{(1+\epsilon)/2}\log{n})$ rounds, w.h.p. Carefully plugging in the right values of $n^{\epsilon}=\tilde{O}(n^{1/3})$ and $n^{\epsilon}=\tilde{O}(n^{1/2})$ then gives the round complexities for triangle detection and listing.

The breakthrough that came after~\cite{Izumi+PODC17} is due to Chang, Pettie, and Zhang~\cite{Chang+SODA19}, who showed triangle listing (and thus also detection) in the \congest model in $\tilde{O}(n^{1/2})$ rounds, w.h.p. In a nutshell, the main methodology of this algorithm has two elements: One element is an algorithm for decomposing the graph into well-connected components which could behave somewhat similarly to the \clique model. The second element is to have the nodes of each component search for triangles that have an edge in the component.

In more detail,~\cite{Chang+SODA19} showed how to compute the following expander decomposition in $O(n^{1/2})$ rounds. This decomposition is a partition of the set of edges of the graph into three sets. In the first set $E_m$, each connected component has minimum degree $n^{1/2}$ and conductance $\Omega(1/\polylog{n})$. The graph induced by the edges in the second set, $E_s$, is such that its arboricity is at most $n^{1/2}$. The third set of remaining edges, $E_r$ is at most a constant fraction of the total number of edges, and the triangle listing algorithm which we discuss in what follows recurses over this remaining set of edges. The algorithm for the decomposition itself is beyond the scope of this survey, and here we only mention that it is rather far from being merely a distributed implementation of its centralized counterpart.\footnote{The actual decomposition result is more general, with a parameter $\delta$ which is tuned here to be $\delta=1/2$ in order to optimize the running time of the triangle listing algorithm that uses it.}

The triangle listing algorithm over the two first sets above then works as follows. Since the arboricity in $E_s$ is bounded by $n^{1/2}$, triangles with at least one edge in this set are listed in a straightforward manner, using an orientation that is deduced by the decomposition algorithm. Then, triangles with an edge in any high-conductance component (i.e., in $E_m$)  are listed by having the nodes of each component mimic a variant of the triangle listing algorithm in the \clique model of Dolev, Lenzen, and Peleg~\cite{Dolev+DISC12}. This mimicking has three aspects. First, it replaces the deterministic partition of~\cite{Dolev+DISC12} with a randomized partition, in order to have a good probability for a good balance of information within the component. Second, the fact that a component has large conductance indeed implies that it has a small mixing time, but this alone is insufficient for efficiently exchanging large amounts of information within the component. To this end, the algorithm makes use of the random-walk based routing techniques of Ghaffari, Kuhn, and Su~\cite{Ghaffari+PODC17} and Ghaffari and Li~\cite{Ghaffari+DISC18}, whose complexities improve as the mixing time decreases. Finally, even with these ingredients, some nodes of a component may have too many edges that touch them that are not inside the component, preventing them from efficiently using their inner-component edges for routing this large amount of information. Thus, some additional edges are added to the set $E_r$ of remaining edges that the decomposition left for recursing over.

The decomposition approach was refined by Chang and Saranurak~\cite{Chang+PODC19}, allowing the complexity of triangle listing to drop to $\tilde{O}(n^{1/3})$ rounds, w.h.p. Since the $\tilde{\Omega}(n^{1/3})$ lower bound for triangle listing in the \clique model directly applies also in \congest, this complexity is near-optimal (up to $\polylog{n}$ factors). At a high level, the decomposition obtained in~\cite{Chang+PODC19} improves upon the one in~\cite{Chang+SODA19} in that it does not have $E_s$ at all (the bounded arboricity part) and in the complexity of obtaining it. In addition, a somewhat modified version of the routing algorithms of~\cite{Ghaffari+PODC17,Ghaffari+DISC18} is introduced, for the triangle listing usage of the decomposition.

An additional algorithm for triangle listing is given by Huang, Pettie, Zhang, and Zhang~\cite{HuangPZZ20}. The complexity of this algorithm, given in terms of the maximum degree $\Delta$,  is $O(\Delta/\log{n}+\log\log\Delta)$ rounds, w.h.p. For $\Delta=\tilde{O}(n^{1/3})$, this is faster compared with the algorithm of~\cite{ChangS20}, while the latter is faster for the larger range of $\Delta$.

Chang and Saranurak~\cite{ChangS20} then show deterministic algorithms for expander decomposition and for expander routing. These yield round complexities of $O(n^{0.58})$ and $n^{2/3}+o(1)$ for deterministic triangle detection and triangle listing, respectively. Building upon this deterministic expander decomposition and routing, Censor-Hillel, Leitersdorf, and Vulakh~\cite{CensorHLV22} showed deterministic triangle listing in $n^{1/3+o(1)}$ rounds. This complexity nearly resolves the question of deterministic triangle listing, but is slightly above the lower bound of $\tilde{\Omega}(n^{1/3})$ rounds, due to the small $n^{o(1)}$ factor that arises from the deterministic expander routing algorithm. Then, Chang, Huang, and Su~\cite{ChangHS24} were able to improve upon deterministic routing, such that plugging it into the framework of~\cite{CensorHLV22}, yields triangle listing in $\tilde{O}(n^{1/3})$ rounds. This leaves only a $\polylog{n}$ gap between the upper and lower bounds for deterministic triangle listing in \congest.

%, leaving the following open question.
%\begin{open}
%\label{open:trianlge-congest-upper-listing}
%Can deterministic expander routing be done in $\polylog{n}$ rounds to yield optimal deterministic triangle listing in the \congest model, or is there a provable gap between the randomized and deterministic complexities for this task?
%\end{open}

The above are triangle listing algorithms so they clearly also solve the detection variant. However, as opposed to triangle listing, for triangle detection the currently known lower bound is still very far from the upper bound. What we do know is the following. Abboud, Censor-Hillel, Khoury, and Lenzen~\cite{AbboudCHKL} showed that triangle detection cannot be solved in the \congest model within a single round by a deterministic algorithm. Specifically, they showed that any single-round algorithm requires a bandwidth of $\Delta\log{n}$ bits. Fischer, Gonen, Kuhn, and Oshman~\cite{Fischer+SPAA18} showed that this also holds for randomized algorithms, by showing that randomized single-round algorithms require a bandwidth of $\Delta$ bits. Both works also addressed the round complexity of 1-bit bandwidth algorithms, with a lower bound of $\Omega(\log^{*}{n})$ rounds given in~\cite{AbboudCHKL}, which was improved to  $\Omega(\log{n})$ rounds in~\cite{Fischer+SPAA18}.\footnote{The function $\log^{*}{n}$ counts the number of times the logarithm function needs to be applied starting from $n$ until the value drops to at most 1.} Later, Assadi and Sundaresan~\cite{AssadiS25} showed an $\Omega(\log\log{n})$ lower bound for triangle detection using new information theoretic arguments.

We will later (Section~\ref{sec:cycles}) describe some lower bounds for finding other subgraphs in the \congest model which are based on reductions from 2-party communication complexity problems, and there we will see why these techniques do not give any meaningful lower bound for triangles. Moreover, in the spirit of the aforementioned argument of Drucker, Kuhn, and Oshman~\cite{Drucker+PODC14}, Eden, Fiat, Fischer, Kuhn, and Oshman~\cite{Eden+DISC19} showed that any polynomial lower bound of $\Omega(n^{\alpha})$ for some constant $\alpha$ for triangle detection in \congest would imply major breakthroughs in circuit complexity. This leaves us with another curious gap in our knowledge of triangle finding in this model.

\begin{open}
\label{open:trianlge-congest-upper}
What is the complexity of triangle detection in the \congest model (randomized and deterministic)?
\end{open}

\section{Finding Larger Cliques}
\label{sec:clique}
\textbf{The \clique model:} The deterministic triangle listing algorithm of~\cite{Dolev+DISC12} for the \clique model can be easily generalized to list larger cliques. To find cliques of size $p$ for some integer $p\geq 3$, the set of nodes is partitioned into $n^{1/p}$ sets. Each node is assigned a $p$-tuple of sets and learns all edges between any two of its sets. A similar argument to that of triangles shows that indeed all $p$-cliques are listed by this algorithm, and that its round complexity is
 $O(n^{1-2/p}/\log{n})$. In fact, it is easy to see that this algorithm lists all instances of any subgraph $H$ of $p$ nodes.

\sloppy{
This complexity is optimal, due to a lower bound of $\tilde{\Omega}(n^{1-2/p})$ rounds by Fischer, Gonen, Kuhn, and Oshman~\cite{Fischer+SPAA18}, which generalizes the aforementioned lower bound for triangle listing of~\cite{Pandurangan+SPAA18} and~\cite{Izumi+PODC17}, using additional machinery. In the \bcc model, Drucker, Kuhn, and Oshman~\cite{Drucker+PODC14} show that $\Omega(n/\log{n})$ rounds are needed for $p$-cliques, even for the stronger detection variant, for almost all values of $p\geq 4$ (as long as $p\leq (1-\epsilon)n$ for some constant $\epsilon>0$).
 }

For sparse graphs, Censor-Hillel, Le Gall, and Leitersdorf~\cite{Censor+PODC20} show that listing can be complete within $\tilde{O}(m/n^{1+2/p}+1) $ rounds, for $p \geq 3$. This follows from their \congest approach which is discussed below, and is tight up to polylogarithmic factors using the lower bound technique of~\cite{Fischer+SPAA18,Pandurangan+SPAA18, Izumi+PODC17}. This complexity was later obtained in a deterministic manner by Censor-Hillel, Fischer, Gonen, Le Gall, Leitersdorf, and Oshman~\cite{Censor-HillelFGGLO20}.

As opposed to the listing variant, for $p$-clique detection the aforementioned hardness of obtaining lower bounds in the \clique model by~\cite{Drucker+PODC14} kicks in, and we do not know any non-constant lower bound (or any larger than 1 lower bound, for that matter), leaving the complexity of $p$-clique detection open for $p>4$.

\begin{open}
\label{open:clique-clique}
What is the complexity of $p$-clique detection in the \clique model (randomized and deterministic), for $p\geq 4$?
\end{open}

~\\\textbf{The \congest model:} Algorithms for finding larger cliques in the \congest model, as in the case of triangles, are also based on the conductance decomposition algorithms of Chang, Pettie, and Zhang~\cite{Chang+SODA19} and of Chang and Saranurak~\cite{Chang+PODC19}. The first sublinear algorithms for larger cliques in \congest are due to Eden, Fiat, Fischer, Kuhn, and Oshman~\cite{Eden+DISC19}. They show that $4$-cliques can be listed in $\tilde{O}(n^{5/6+o(1)})$ rounds and that $5$-cliques can be listed in $\tilde{O}(n^{73/75+o(1)})$ rounds, w.h.p. These listing algorithms are more involved compared with the triangle listing algorithm, due to the need to handle, for example, a $4$-clique with one edge within a certain component and another edge within a different component, which imposes a complex challenge that becomes even worse as $p$ grows.

Following~\cite{Eden+DISC19}, the work of Censor-Hillel, Le Gall, and Leitersdorf~\cite{Censor+PODC20} provided algorithms for listing $p$-cliques in $\tilde{O}(n^{p/(p+2)})$ rounds, w.h.p., for all $p\geq 4$ except for $p=5$. For $p=5$, this work gave an algorithm which completes in $\tilde{O}(n^{3/4+o(1)})$ rounds, w.h.p. The main approach of these algorithms is iterating over the decomposition in a way that balances the minimum degree and the arboricity thresholds. Within each cluster, sparsity-aware listing helps speeding up the computation. While these algorithms improved upon the state-of-the-art for $p=4,5$ and were the first sublinear algorithms for $p\geq 6$, they fell short of the~\cite{Fischer+SPAA18} lower bound of $\tilde{\Omega}(n^{1-2/p})$ rounds. The question of whether clique listing in the \congest model is as easy as, or harder than, its \clique counterpart, was recently answered by Censor-Hillel, Chang, Le Gall, and Leitersdorf~\cite{Censor+SODA21}, using the newer conductance decomposition of Chang and Saranurak~\cite{Chang+PODC19} and additional mechanisms for optimal sparsity-aware listing and for efficient transmission of edges in the clusters. Deterministic algorithms for $p$-clique listing were given by Censor-Hillel, Leitersdorf, and Vulakh~\cite{CensorHLV22}, with round complexities of $n^{1-2/p+o(1)}$. As is the case for triangles, these complexities are optimal up to the small $n^{o(1)}$ factor that arises from the expander routing procedure of Chang and Saranurak~\cite{ChangS20}. Later, the aforementioned improved deterministic expander routing of Chang, Huang, and Su~\cite{ChangHS24} was plugged into the framework of~\cite{CensorHLV22} to obtain a round complexity of $\tilde{O}(n^{1-2/p})$, leaving only a $\polylog{n}$ gap compared with the lower bound.

For $p=4$, the tight listing of~\cite{Censor+SODA21} also implies that $4$-clique detection is not easier than its listing counterpart in the \congest model. This is due to a lower bound of $\tilde{\Omega}(n^{1/2})$ for $4$-clique detection in \congest, by Czumaj and Konrad~\cite{Czumaj+DISC18}. The lower bound of~\cite{Czumaj+DISC18} is more general, and says that detecting $p$-cliques in the \congest model requires $\tilde{\Omega}(n^{1/2}/p)$ rounds for $p\leq n^{1/2}$, and $\tilde{\Omega}(n/p)$ rounds for $p\geq n^{1/2}$. Thus, for values of $p$ larger than $4$, there is still a gap between detection and listing.

\begin{open}
\label{open:clique-congest}
What is the complexity of $p$-clique detection in the \congest model (randomized and deterministic), for $p> 4$?
\end{open}

The lower bound of~\cite{Czumaj+DISC18} uses a reduction from 2-party set disjointness. The same work also shows that listing can be done sufficiently fast by the two parties, which shows that if the detection problems are harder, a different lower bound technique must be used.

\section{Finding Larger Cycles}
\label{sec:cycles}
\textbf{The \clique model:}
Since the algorithm of~\cite{Dolev+DISC12} easily lists any subgraph of $p$ nodes, we have that, in particular, $p$-cycles can be listed in the \clique model within $O(n^{1-2/p})$ rounds (deterministically).

The detection variant was then addressed by Censor-Hillel, Kaski, Korhonen, Lenzen, Paz, and Suomela~\cite{Censor-Hillel+DC19}, who show a deterministic constant-round algorithm for detecting $4$-cycles. In addition, they show an $O(n^{0.158})$-round algorithm for detecting $p$-cycles, for any constant $p$. More accurately, the complexity is $2^{O(p)}n^{0.158}$ rounds. This is a randomized algorithm that relies on the color coding technique of Alon, Yuster, and Zwick~\cite{AlonYZ95}, in order to avoid too much congestion that would be caused by collecting all possible candidates for cycle nodes. More recently, Censor-Hillel, Fischer, Gonen, Le Gall, Leitersdorf, and Oshman~\cite{Censor-HillelFGGLO20} showed that $2p$-cycles can be detected in $O(1)$ rounds for any $p$, using a technique which also allows fast girth approximation. This is, notably, a deterministic algorithm. For odd values of $p$, it is still not known whether there is a constant-round $p$-cycle detection algorithm.

\begin{open}
\label{open:oddcycles-cc}
What is the complexity of $p$-cycle detection in the \clique model (randomized and deterministic), for odd values of $p$?
\end{open}

For the parametrized version, in which the complexity depends on the number of $p$-cycles in the graph, the aforementioned work of Censor-Hillel, Even, and Vassilevska Williams~\cite{Censor-HillelEW24} gives a complexity of $\tilde{O}(p^{O(p)}\cdot n^{0.1567}/(t^{\frac{0.4617}{p-0.82408}}+1))$ rounds.

~\\\textbf{The \congest model:}
For $4$-cycle detection, Drucker, Kuhn, and Oshman~\cite{Drucker+PODC14} provide a tight complexity of $\tilde{\Theta}(n^{1/2})$, by providing both an algorithm and a lower bound. They also show that for any odd value of $p$, detecting $p$-cycles requires $\tilde{\Omega}(n)$ rounds. The upper bound is obtained by setting a threshold of $T=2n^{1/2}$, and defining heavy nodes as nodes with at least $T$ neighbors. First, all non-heavy nodes send all their neighbors to all their neighbors, in $T$ rounds. This detects $4$-cycles which have at least 2 non-neighboring non-heavy nodes. Then, a heavy node $v$ with more than $T$ heavy neighbors reports that there exists a $4$-cycle. Finally, a heavy node $v$ with at most $T$ heavy neighbors sends its heavy neighbors to all of its neighbors, again in $T$ rounds. The reason for which too many heavy neighbors imply a $4$-cycle is because this means that the total number of neighbors of heavy neighbors is at least $T^2$, which is more than $2n$, and so there must be a node other than $v$ which is connected to two of the neighbors of $v$, and hence we have a $4$-cycle. Otherwise, if $v$ sends its heavy neighbors to all of its neighbors, then any $4$-cycle with two neighboring nodes that are heavy is detected by one of its nodes (by a simple case analysis).

The matching lower bound for $4$-cycles is a good point of reference for understanding lower bounds for the \congest model that rely on reductions from 2-party communication complexity. It relies on the fact that there is a $4$-cycle-free graph $G$ on $n$ nodes with $\Theta(n^{3/2})$ edges, due to Erd{\"o}s~\cite{Erdos38}, whose work implies that this is the Tur{\'a}n Number of $4$-cycles~\cite{Turan41}. The setting is as follows. Each of two players, Alice and Bob, has an input string of $k$ bits, $x=(x_1,\dots, x_k)$ and $y=(y_1,\dots, y_k)$, respectively. By communicating with each other, the players  need to compute the set disjointness function $Disj(x,y)$, whose value is 1 if and only if the input strings represent disjoint subsets of the set $\{1,\dots,k\}$, that is, if and only if there is no index $1\leq i \leq k$ such that $x_i=y_i=1$. The 2-party set disjointness problem is known to require exchanging $\Omega(k)$ bits, even by randomized protocols due to Kalyanasundaram and Schnitger~\cite{KalyanasundaramS92}, Razborov~\cite{Razborov90}, and Bar{-}Yossef, Jayram, Kumar, and Sivakumar~\cite{Bar-YossefJKS04}. The reduction to distributed detection of $4$-cycles is as follows. Each of the two players takes a subgraph of the $4$-cycle free graph $G$ according to their respective input. That is, the edges of $G$ are mapped to the set $\{1,\dots,k\}$, with a value of $n$ for which $k=\Theta(n^{3/2})$. Each player imagines a subgraph of $G$ that has an edge for any index in which their input is 1. The players then imagine that their two subgraphs are connect by a perfect matching: each node in Alice's subgraph is connected to the respective node in Bob's subgraph (the nodes of both subgraphs are the nodes of $G$). Now, it is easy to see that the combined graph contains a $4$-cycle if and only if the input strings $x,y$ are disjoint, as the only $4$-cycles that can occur are those that have two edges from the perfect matching, and two edges that represent the same index in the bit strings of the players. Thus, if Alice and Bob can simulate a distributed algorithm for $4$-cycle detection, they solve set disjointness. They simulate a given algorithm as follows. Any message that the algorithm sends between two nodes that are in the same subgraph (either that of Alice of that of Bob) can be internally simulated by the respective player. The only messages that need to be explicitly sent between the two players are those that are sent along the edges of the perfect matching. There are $n$ such edges, and so simulating a round of the detection algorithms costs $O(n\log(n))$ bits of communication between the two players. Since $\Omega(k)=\Omega(n^{3/2})$ is a lower bound on the total number of bits that need to be exchanged for solving set disjointness, we get that the distributed $4$-cycle detection algorithm must consist of at least $\Omega(k/n\log(n))=\Omega(n^{3/2}/n\log(n))=\tilde{\Omega}(n^{1/2})$ rounds.

\textbf{A remark on triangle detection:} While this approach for reductions which imply \congest lower bounds is heavily used in the literature, it is doomed to fail for triangle detection. The reason is that any triangle in a graph has at least one player which knows at least two of its nodes and thus all of its edges, which nullifies any attempt for a lower bound constructions, regardless of the 2-party problem or the graph construction.

~\\
Korhonen and Rybicki~\cite{KorhonenR17} then showed that the $\tilde{\Omega}(n^{1/2})$ lower bound for $4$-cycles can be extended and applies to $p$-cycles for all even values of $p$. Notably, in the spirit of the results of~\cite{Drucker+PODC14}, it is shown in Censor-Hillel, Fischer, Gonen, Oshman, Le Gall, and Leitersdorf~\cite{Censor-HillelFGGLO20} that going above this lower bound for $p=6$ would imply new lower bounds in circuit complexity, which are considered hard to obtain. The reason that this barrier for lower bounds works in the \congest model is because it is shown that it is sufficient to consider high-conductance clusters, and that these can simulate circuits in a similar manner to the \clique model.

On the upper bound side, Korhonen and Rybicki~\cite{KorhonenR17} showed that $p$-cycle detection for any constant $p$ can be done in a linear in $n$ number of rounds. For odd values of $p$, this is optimal due to the above lower bound. They also showed that this can be done faster for degenerate graphs. Fischer, Gonen, Kuhn, and Oshman~\cite{Fischer+SPAA18} showed how to detect $2p$-cycles within $O(n^{1-1/p(p-1)})$ rounds, for $p\geq 2$. This was subsequently improved by Eden, Fiat, Fischer, Kuhn, and Oshman~\cite{Eden+DISC19}, who showed how to detect $2p$-cycles in $\tilde{O}_p(n^{1-2/(p^2-p+2)})$ rounds for odd $p \geq 3$, and in $\tilde{O}_p(n^{1-2/(p^2-2p+4)})$ rounds for even $p \geq 4$. They also show that as opposed to the case of $4$-cliques, the listing variant of $4$-cycles is harder than its detection counterpart, requiring $\tilde{\Omega}(n)$ rounds. For $p=3,4,5$, Censor-Hillel, Fischer, Gonen, Oshman, Le Gall, and Leitersdorf~\cite{Censor-HillelFGGLO20} then showed improved algorithms (randomized) for detecting $2p$-cycles, which completes within $\tilde{O}(n^{1-1/p})$ rounds. Later, Fraigniaud, Luce, and Todinca~\cite{FraigniaudLT23} showed that the technique of~\cite{Censor-HillelFGGLO20} provably cannot go beyond these values of $p$. Following this, Fraigniaud, Luce, Magniez, and Todinca~\cite{FraigniaudLMT24} showed a major enhancement of this technique which does apply to larger values of $p$, yielding a complexity of $\tilde{O}(n^{1-1/p})$ rounds for detecting $2p$-cycles for any value of $p$. Later, they showed how to obtain this complexity without the use of randomization \cite{FraigniaudLMT25}.

Still, the upper bounds here are above the respective $\tilde{\Omega}(n^{1/2})$ lower bound.

\begin{open}
\label{open:oddcycles-congest}
What is the complexity of $p$-cycle detection in the \congest model (randomized and deterministic), for even value of $p\geq 6$?
\end{open}

Finding \emph{induced} cycles is a different task, as it is insufficient to have edges that form a cycle, but rather also requires no other edges between nodes of the cycle. This question was addressed by Le Gall and Miyamoto~\cite{LeGallM21}, who showed a lower bound of $\tilde{\Omega}(n)$ rounds for detecting induced $p$-cycles for any $p\geq 4$. This implies that detection of induced cycles is strictly harder than the non-induced case. This bound is tight for $p=4$, and this work also shows that obtaining a higher lower bound for $p=5,6,7$ cannot be done using the standard framework of reductions from 2-party communication complexity problems. For larger values of $p\geq 8$, a lower bound of $\tilde{\Omega}(n^{2-\Theta(1/p)})$ is given. In addition, this work studies the problem of finding induced diamonds, which are $4$-cycles with exactly one additional edge.

\section{Additional Variants of Distributed Subgraph Finding}
\label{sec:additional}

\textbf{Finding Other Subgraphs:}
The literature has also been studying additional subgraphs, apart from cliques and cycles.

Drucker, Kuhn, and Oshman~\cite{Drucker+PODC14} study the complexity of finding trees and complete bipartite subgraphs in the \bcc model. Korhonen and Rybicki~\cite{KorhonenR17}  show that trees can be detected in $O(1)$ rounds in the \bcongest model. Nikabadi and Korhonen~\cite{NikabadiK21} show algorithms for induced subgraphs, and lower bounds for induced cycles were given by Le Gall and Miyamoto~\cite{LeGallM21}.

On the lower bound front, Gonen and Oshman \cite{Gonen+OPODIS17} showed lower bounds for finding subgraphs that are created from smaller ones using certain allowed operations. Fischer, Gonen, Kuhn, and Oshman~\cite{Fischer+SPAA18} showed that for any $p\geq 4$, there exists a subgraph $H$ of size $p$ such that $H$ detection requires $\tilde{\Omega}(n^{2-\Theta(1/p)})$ rounds.  Eden, Fiat, Fischer, Kuhn, and Oshman~\cite{Eden+DISC19} showed that this complexity is roughly the right one, by providing an upper bound of $\tilde{O}(n^{2-2/(3p+1)}+o(1))$ rounds. In particular, their result shows that there does not exist a constant-sized subgraph for which detection requires a truly quadratic number of rounds.

~\\\textbf{Subgraph Freeness Testing:} The variant of testing for $H$-freeness has also been studied in the distributed setting, initially introduced by Brakerski and Patt-Shamir~\cite{BrakerskiP11}.

The definition of distributed testing for $H$-freeness requires that if there is no instance of $H$ in the graph then all the nodes output \texttt{true}, but instead of requiring at least one node to output \texttt{false} in case there is an instance of $H$ as would be an analog to the detection problem, testing only requires at least one node to output \texttt{false} in case the graph is far from being $H$-free. Here, being $\epsilon$-far from having a property is the same as its standard definition by Goldreich, Goldwasser, and Ron~\cite{GoldreichGR98}, and means that no matter which $\epsilon$-fraction of the graph is changed (removed, in the case of $H$-freeness), the resulting graph must still have a copy of $H$. For triangles and other small subgraphs, this typically means that there are many instances of $H$ in a graph that is far from being $H$-free.

For testing triangle-freeness, Censor-Hillel, Fischer, Schwartzman, and Vasudev~\cite{Censor-HillelFSV19} showed a $O(1/\epsilon^2)$-round algorithm. This was later improved by Even, Fischer, Fraigniaud, Gonen, Levi, Medina, Montealegre, Olivetti, Oshman, Rapaport, and Todinca~\cite{Even+DISC17} and Fraigniaud and Olivetti~\cite{FraigniaudO17} to a complexity of $O(1/\epsilon)$ rounds. These, along with the work of  Fraigniaud, Rapaport, Salo, and Todinca~\cite{FraigniaudRST16} also show fast testing for larger cliques, cycles, and additional subgraphs.

Somewhat related is the work of Censor-Hillel and Khoury~\cite{CensorHillelK24}, which showed upper and lower bounds for computing and approximating the distance of a graph to being triangle-free.

~\\\textbf{Subgraph Finding in Dynamic Networks:} Subgraph finding has also been studied from the perspective of dynamic networks.
Bonne and Censor-Hillel~\cite{BonneC19} characterize the bandwidth that is required for various clique finding problems by algorithms that work in a dynamic setting and must produce the correct answer immediately at the end of the round in which a topology change occurs.

Subgraph finding in a harsh dynamic setting in which the number of topology changes per round is unlimited, was studied by Censor-Hillel, Kolobov, and Schwartzman~\cite{CHKS-dyn}, who showed upper and lower bounds for small subgraphs.

~\\\textbf{Subgraph Finding in Quantum Networks:}
Izumi, Le Gall, and Magniez~\cite{Izumi+STACS20} have shown that in a quantum \congest model, triangle detection can be solved within $\tilde{O}(n^{1/4})$ rounds, using a distributed version developed by Izumi and Le Gall in~\cite{Izumi+PODC19} of a \emph{Grover Search}~\cite{Grover96}. Later, Censor-Hillel, Fischer, Le Gall, Leitersdorf, and Oshman showed quantum triangle detection in $\tilde{O}(n^{1/4})$ rounds and clique detection for larger cliques, using nested Grover searchers~\cite{Censor-HillelFLGLO22}. Later, van Apeldoorn and de Vos~\cite{ApeldoornV22} have shown algorithms for cycle detection and girth computation in the quantum \congest model. The work of Fraigniuad, Luce, Magniez, and Todinca~\cite{FraigniaudLMT24} further provides faster algorithms for cycle detection in the quantum \congest model.

The aforementioned cycle detection algorithm of~\cite{Censor-HillelEW24} in the \clique model, gives a triangle detection algorithm in the quantum \clique model. Intuition of why this currently does not extend to larger cycles is given therein. 

We note that the listing variant for triangles cannot be improved using quantum tools, since it is obtained through information theoretic arguments, which apply to this setting as well.

~\\\textbf{Acknowledgements.} This project has received funding from the European Union’s Horizon 2020 research and innovation programme under grant agreement no. 755839. The author thanks Francois Le Gall for clarifications and Dean Leitersdorf for useful feedback on a preliminary version of this survey.

\bibliographystyle{alpha}
\bibliography{References}

\end{document}